\documentclass[aps,pre,showpacs,twocolumn,superscriptaddress,longbibliography,amsmath,amssymb]{revtex4-2}
\usepackage[pdftex]{graphicx}
\usepackage{color}
\usepackage{xspace}
\usepackage[pdftex,bookmarks,colorlinks,breaklinks]{hyperref} 
\hypersetup{linkcolor=red,citecolor=blue,filecolor=dullmagenta,urlcolor=blue} 
\hypersetup{pdftoolbar=true,pdfpagemode=UseNone,pdfstartview=FitH, colorlinks=true,extension=}

\newcommand{\Tr}{\mathrm{Tr\,}}
\newcommand{\Tjunctions}{{\fontfamily{cmss}\selectfont T} junctions\xspace}
\begin{document}
\title{A spectral duality in graphs and microwave networks}
\author{Tobias Hofmann}
\affiliation{Fachbereich Physik, Philipps-Universit\"{a}t Marburg, 35032 Marburg, Germany}
\author{Junjie Lu}
\affiliation{Institut de Physique de Nice, CNRS, Universit\'{e} C\^{o}te d'Azur, 06108 Nice, France, European Union}
\author{Ulrich Kuhl}
\email{ulrich.kuhl@univ-cotedazur.fr}
\affiliation{Institut de Physique de Nice, CNRS, Universit\'{e} C\^{o}te d'Azur, 06108 Nice, France, European Union}
\affiliation{Fachbereich Physik, Philipps-Universit\"{a}t Marburg, 35032 Marburg, Germany}
\author{Hans-J\"urgen~St\"{o}ckmann}
\email{stoeckmann@physik.uni-marburg.de}
\affiliation{Fachbereich Physik, Philipps-Universit\"{a}t Marburg, 35032 Marburg, Germany}

\date{\today}

\begin{abstract}
Quantum graphs and their experimental counterparts, microwave networks, are ideally suited to study the spectral statistics of chaotic systems.
The graph spectrum is obtained from the zeros of a secular determinant derived from energy and charge conservation.
Depending on the boundary conditions at the vertices, there are Neumann and Dirichlet graphs.
The first ones are realized in experiments, since the standard junctions connecting the bonds obey Neumann boundary conditions due to current conservation.
On average, the corresponding Neumann and Dirichlet eigenvalues alternate as a function of the wave number, with the consequence that the Neumann spectrum is described by random matrix theory only locally, but adopts features of the interlacing Dirichlet spectrum for long-range correlations.
Another spectral interlacing is found for the Green's function, which in contrast to the secular determinant is experimentally accessible.
This is illustrated by microwave studies and numerics.
\end{abstract}


\maketitle

\section{Introduction}

Quantum graphs, formed by connected networks of bonds and vertices, are ideally suited to study questions coming from quantum chaos and random matrix theory (RMT).
In closed quantum graphs the main interest has been focused on the statistical properties of the spectra.
Most studies in this respect were motivated by the famous conjecture by Bohigas, Giannoni, and Schmit (BGS) that the universal features of the spectra of chaotic systems should be described by RMT \cite{boh84b}.
Using supersymmetry techniques, Gnutzmann and Altland \cite{gnu04b} proved the BGS conjecture for the two-point correlation function for graphs with incommensurate bond lengths.
Their result was generalized to all correlation functions by Pluha{\v r} and Weidenm\"{u}ller \cite{plu14}, who furthermore proved the applicability of RMT to the scattering properties of graphs \cite{plu13a}.

Just as for billiard systems \cite{stoe90}, there is a one-to-one correspondence between a quantum graph and the corresponding microwave network, called a microwave graph in the following.
This correspondence has been used in particular by Sirko and co-workers, in numerous experiments to study spectral and scattering properties of microwave graphs (see Ref.~\cite{hul04} as an example).
A specific feature of open graphs is topological resonances corresponding to states existing exclusively within the system and being invisible from the outside \cite{gnu13}.
Specifically designed graphs were used to mimic spin-$\frac{1}{2}$ systems for the first experimental realization of the Gaussian symplectic ensemble \cite{reh16,reh18}, following an idea by Joyner {\it et~al.}~\cite{joy14}.
Most recent applications of graphs have been on non-Weyl graphs \cite{law19b} and in the study of coherent perfect absorption and complex zeros of the scattering matrix of graphs \cite{che20}.

All experiments as well as numerical studies have been performed on graphs with a small number of vertices $V$ typically below 10-20, whereas the above-mentioned proofs of the applicability of RMT hold only for strongly connected graphs in the limit $V\to\infty$.
This point will become important later.

In the microwave studies the graphs are realized in terms of networks formed by cables connected by \Tjunctions at the vertices.
A vector network analyzer measures the reflection at one port attached to the graph, or the transmission from one port to another, if there are more of them.
The total $S$ matrix is available experimentally, including the phases, a unique property of the technique.
The scattering matrix contains all the information needed for the determination of the graph spectrum.
It will be called the Neumann spectrum in the following, since the \Tjunctions at all vertices obey Neumann boundary conditions.

This specification is needed since in the following another spectrum will be of importance, the Dirichlet spectrum, describing a graph where all vertices obey Dirichlet boundary conditions.
This situation corresponds to a totally disintegrated graph with a spectrum being the sum of the spectra of all individual bonds with Dirichlet boundary conditions at both ends.
Both spectra are tightly interlaced with each other, a phenomenon generically found in systems subject to rank-1 perturbations \cite{sim95b}.
Consequences of the change of boundary conditions from Neumann to Dirichlet (and any other situation in between) were discussed in the monograph by Berkolaiko and Kuchment \cite{ber13}.
The implications of interlacing for the spectral statistics in the context of RMT, however, have not been considered so far.

\section{Theory}

\subsection{The graph secular equation system}

For a better understanding of the interplay between the Neumann and the Dirichlet spectrum we have to look somewhat more in detail into the mathematical description of graphs.
In the presentation we follow the work by Kottos and Smilansky \cite{kot99a}.

In the accessible frequency range the cables used in the experiments support only one propagating mode.
The wave fields within the graph have to obey two constraints.
The first one is energy conservation, meaning that at each vertex $n$ there exists a unique potential $\varphi_n$ for all bonds meeting at this vertex.
This condition is automatically met by means of the ansatz
\begin{equation} \label{eq:potential}
  \psi_{nm}(x)=\frac{1}{\sin kl_{nm}}\left[\varphi_n\sin k(l_{nm}-x)+\varphi_m\sin kx\right]
\end{equation}
for the wave within the bond connecting vertices $n$ and $m$, where $x$ is the distance to vertex $n$ and $l_{nm}$ is the length of the bond.
Equation~(\ref{eq:potential}) holds for time reversal graphs, the only ones considered here.

The second constraint is current conservation at each vertex $n$,
\begin{equation} \label{eq:current}
 \sum\limits_m \left.\frac{\mathrm{d}\psi_{nm}(x)}{\mathrm{d}x}\right|_{x=0}=0\,,
\end{equation}
where the sum is over all bonds $m$ meeting at vertex $n$.
Equation~(\ref{eq:current}) holds for Neumann boundary conditions at the vertices.

Plugging the expression (\ref{eq:potential}) into Eq.~(\ref{eq:current}), we obtain an equation system for the potentials,
\begin{equation} \label{eq:hom}
 \sum\limits_{m}h_{nm}\varphi_m=0\,,
\end{equation}
where
\begin{equation} \label{eq:hsec}
  h_{nm}=-\delta_{nm}\sum\limits_{m'}f_{nm'} +g_{nm}\,,
\end{equation}
with
\begin{equation} \label{eq:fg}
  f_{nm}=\cot kl_{nm}\,,\quad g_{nm}=1/\sin k l_{nm}\,,
\end{equation}
if there is a bond connecting vertices $n$ and $m$, and $f_{nm}=g_{nm}=0$ otherwise.
For a dangling bond with Dirichlet boundary condition at the open end the ansatz (\ref{eq:potential}) reduces to
\begin{equation} \label{eq:dang}
  \psi_n(x)=\frac{1}{\sin kl_n}\varphi_n\sin k(l_n-x)
\end{equation}
whence it follows that
\begin{equation} \label{eq:gdang}
  g_n=0
\end{equation}
for dangling bonds.
The end points of dangling bonds will not be counted in the number of vertices, since the boundary condition $\psi_n(l_n)=0$ has already been taken into account by the ansatz (\ref{eq:dang}).

For the homogeneous equation system (\ref{eq:hom}) to have non-trivial solutions the determinant of the matrix $h(k)$ with elements $h_{nm}(k)$ has to vanish,
\begin{equation} \label{eq:det}
  \left| h(k)\right|=0\,.
\end{equation}
The roots $k_n$ of this equation generate the Neumann spectrum of the graph.
On the other hand, $h_{nm}$ becomes singular, whenever $kl_{nm}$ is an integer multiple of $\pi$.
This is exactly the resonance condition for the Dirichlet spectrum belonging to the bond connecting vertices $n$ and $m$.
The Dirichlet spectrum hence appears via the poles of $|h(k)|$.
In the following all lengths will be assumed to be incommensurable to avoid degeneracies of the Dirichlet spectrum, one of the necessary ingredients  to obtain a spectrum statistically described by RMT.

Since the distance between successive $k$ eigenvalues on a bond of length $l_i$ is $\Delta k=\pi/l_i$, each bond contributes with $\rho^D_i=l_i/\pi$ to the density of states of the Dirichlet spectrum.
Its total density of states hence is $\rho^D=\sum_i \rho^D_i= l_\mathrm{tot}/\pi$, where $l_\mathrm{tot}=\sum_il_i$ is the total length of the graph.
However, according to Weyl's law this is identical to the mean density of states of the Neumann spectrum \cite{kot99a}.
Hence both Neumann and Dirichlet spectra have the same mean density of states.

\subsection{The graph Green function}

For an experimental study of the spectral properties, the graph has to be opened by attaching one or more open bonds.
Let us hence assume that there are  open bonds $ n=1,\dots, N$ attached at vertices $1,\dots, N$.
The field within the bonds may be written as the superposition of two waves propagating in opposite directions,
\begin{equation} \label{eq:psi}
  \psi_n(x)=a_ne^{-ikx}-b_ne^{ikx}\,, \quad n=1,\dots, N
\end{equation}
where $x$ is the distance to the vertex, and $a_n$ and $b_n$ are the amplitudes of the waves propagating towards and away from the vertex, respectively.
The definition~(\ref{eq:psi}) corresponds to the convention applied in microwave technology.
It is also in accordance with definitions applied in the context of quantum dots \cite{bee96} and nuclear physics \cite{guh98}.
In quantum graphs \cite{kot99a} another definition is in use, where in contrast to Eq.~(\ref{eq:psi}) both wave components come along with a positive sign.

\begin{figure}
	\includegraphics[width=\columnwidth]{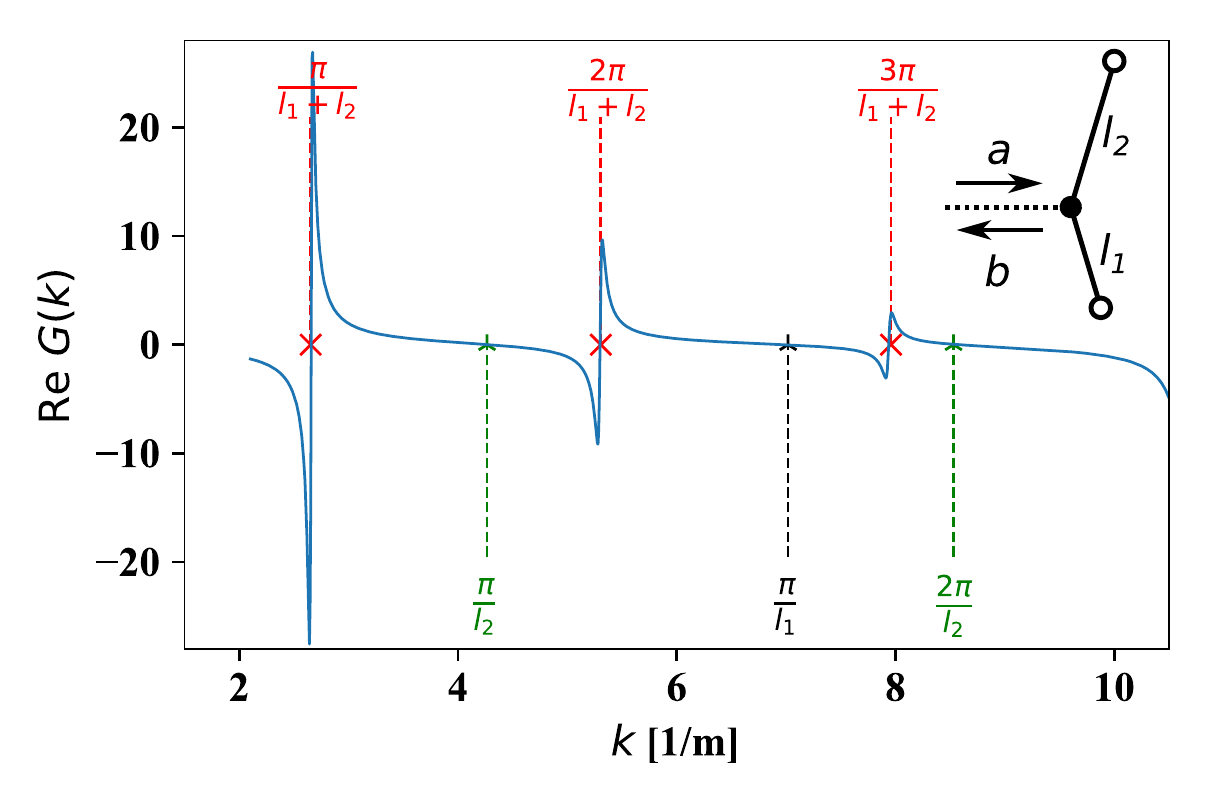}\\
	\caption{\label{fig:fig1}
		Plot of $\mathrm{Re}(G)$ as a function of $k=2\pi\nu/c$ for the graph shown in the inset as obtained from a microwave reflection measurement ($l_{1,2}/\mathrm{m}= 0.448, 0.736$).
		The Neumann spectrum is formed by the poles $k_n$ and the Dirichlet spectrum by the zeros $k^D_n$ of $\mathrm{Re}(G)$.
		Due to absorption, all poles are converted into dispersion-like resonances.
		In all figures Neumann vertices are depicted by filled circles and Dirichlet ones by open circles.}
\end{figure}

The two constraints, energy and current conservation, yield, for vertices $1,\dots, n$,
\begin{equation} \label{eq:potcurr}
  \begin{array}{rcl}
    \varphi_k &=& a_k-b_k \\
    \sum\limits_m h_{km}\varphi_m&=&i(a_k+b_k)
  \end{array}\,,\quad 
\end{equation}
for $k=1,\dots, n\,$.
The equation system (\ref{eq:hom}) now has become inhomogeneous,
\begin{equation} \label{eq:inhom}
  h\varphi=i(a+b)
\end{equation}
where $a=(a_0,\dots, a_n,0,\dots)^T$ and $b=(b_0,\dots, b_n,0,\dots)^T$.
It follows that
\begin{equation} \label{eq:inhom1}
  \varphi=i h^{-1}(a+b)\,.
\end{equation}
At the coupling vertices $1,\dots, N$ the $\varphi_k$ are fixed by the constraints (\ref{eq:potcurr}), whence it follows that
\begin{equation} \label{eq:G}
  a-b=iG(a+b)\,,
\end{equation}
where $G$ is the matrix with elements
\begin{equation} \label{eq:G1}
   G_{kl}=(h^{-1})_{kl}\,, \quad k,l=1, \dots, N\,.
\end{equation}

Incoming and outgoing amplitudes are connected via the scattering matrix $S$,
\begin{equation} \label{eq:S}
  b=Sa\,.
\end{equation}
Equation (\ref{eq:G}) yields for the scattering matrix
\begin{equation} \label{eq:S1}
	S=\frac{1-iG}{1+iG}\,,
\end{equation}
A slightly different expression for the scattering matrix is well known from quantum dots \cite{bee96} and nuclear physics \cite{guh98},
\begin{equation} \label{eq:S2}
	S=\frac{1-iW^\dag G W}{1+iW^\dag G W}\,,
\end{equation}
where $G$ is the Green's function, and the components $W_k$ of vector $W$ specify the coupling strengths to the open channels.
For an ideal coupling, as provided by the \Tjunctions, $W_k=1$ holds for all channels.
The $W_k$ hence do not appear in the present case.

The above derivation of the connection between the scattering matrix and $h$ matrix of the graph is more or less equivalent to the one presented in Refs.~\cite{kot99a,kot04}, which yield, however, the negative of expression (\ref{eq:S1}) for $S$, a consequence of the differing sign convention in Eq.~(\ref{eq:psi}).

\section{Experiment and numerics}

\subsection{Interlacing features of the experimental Green function}

Our microwave equipment allows us to take spectra from 50\,MHz to 20\,GHz.
The networks are constructed from standard microwave coaxial cables connected by \Tjunctions.
In the experimentally accessible frequency range, each cable supports only one propagating mode.
A vector network analyzer (VNA) measures the reflection from one open cable attached to the graph or the transmission from one cable to another one.
Microwave technology is subject to the 50 $\Omega$ convention meaning an ideal matching of the cables connected to the VNA.
Each attached cable hence means an open channel with no reflections from the end.
In the present work only one attached cable has been used; hence the scattering matrix reduces just to a phase factor if absorption is ignored, $S=e^{i\alpha}$, and $G=G_{11}= -\tan(\alpha/2)$ is just a number.

Since $G$, and not $S$, is the main quantity of interest, we present our experimental results in the following in terms of
\begin{equation} \label{eq:G}
  G=-i\frac{1-S}{1+S}\,,
\end{equation}
holding for one open channel.
Mathematically, the conversion from the expression (\ref{eq:S1}) to the expression (\ref{eq:G}) is trivial; experimentally it is not.
Usually, even after a careful calibration there remains a global phase drift of the $S$ matrix of typically about 0.2$\pi$/GHz, which has to be removed before the conversion (\ref{eq:G}) can be performed.

As an illustrative toy example we present experimental results for the graph shown in Fig.~\ref{fig:fig1}.
It consists of just two dangling bonds of lengths $l_1$ and $l_2$ terminated by short ends, corresponding to Dirichlet boundary conditions.
For this case we obtain
\begin{equation} \label{eq:Gtoy}
  G=-\frac{\sin kl_1 \sin kl_2}{\sin k(l_1+l_2)}\,.
\end{equation}
In the presence of absorption, always the case in the experiment, $k$ has
to be replaced by $k+i\lambda$.

For the toy graph there is only one vertex, the secular matrix $h$ becomes just a number, and Eq.~(\ref{eq:G1}) simplifies to $G=h^{-1}$.
The zeros of $h$ are thus converted into poles of $G$ and vice versa.
Both Eq.~(\ref{eq:Gtoy}) and Fig.~\ref{fig:fig1} nicely illustrate these features: The Neumann spectrum, formed by the zeros of $h$, at $k_n=n\pi/(l_1+l_2)$, shows up at the poles of $G$, and the Dirichlet spectrum, formed by the poles of $h$, at $k^D_{1n}=n\pi/l_1$ and $k^D_{2n}=n\pi/l_2$, in the zeros of $G$.
Here one has to keep in mind that a vertex with Neumann boundary conditions and connected to only two bonds may just be removed without changing the spectrum, an immediate consequence of current conservation.

For larger graphs the situation is somewhat more complicated.
Now $G=G_{11}$ is obtained from the matrix inverse of $|h|$,
\begin{equation} \label{eq:G2}
  G=(h^{-1})_{11}=\left|h_{11}\right|/\,|h|\,,
\end{equation}
where $h_{11}$ is the matrix obtained from $h$ by removing the first row and first column.

\begin{figure}
\includegraphics[width=\columnwidth]{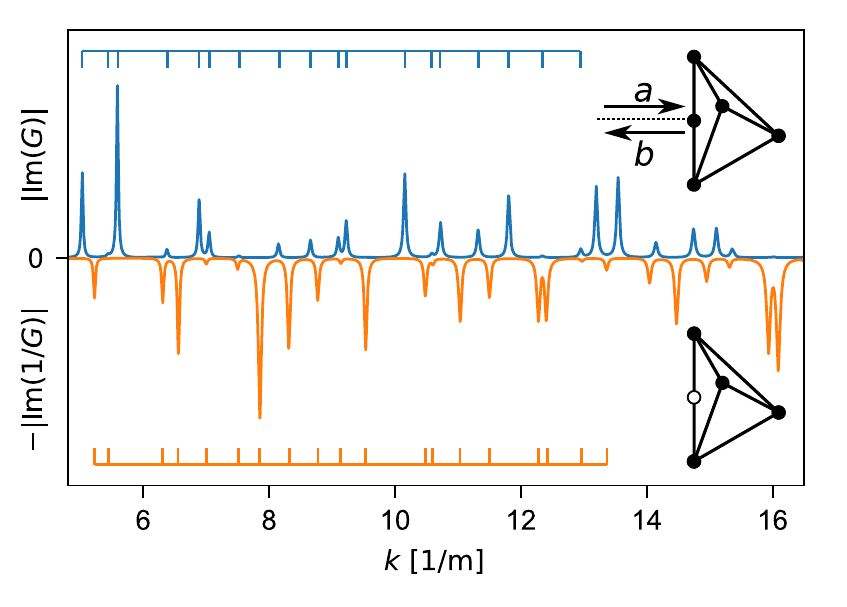}
\caption{\label{fig:fig2}
	Experimental $|\mathrm{Im}(G)|$ (blue) and $-|\mathrm{Im}(G^{-1})|$, (orange) for the upper tetrahedral graph shown on the right ($l_{1,\dots, 7}/\mathrm{m}= 0.376, 0.440, 0.786, 0.870, 0.952, 1.593, 1.754$).
	The combs denote the calculated spectra for the two tetrahedrons depicted on the right.}
\end{figure}

Since the determinant of $h$ appears in the denominator, the Neumann spectrum still is made up of the poles of $G$, but now what is the meaning of the zeros?
This question is answered by the following reasoning:
$G=0$ means, according to Eq.~(\ref{eq:S1}), $S=1$ at the coupling vertex.
The waves are hence totally reflected at the entrance.
But this can only happen, if there is a resonance within the graph with zero amplitude at the coupling point.

{\em Hence the zeros of $G$ represent the spectrum of the graph obtained from the original one by changing the boundary conditions at the coupling point from Neumann to Dirichlet, or, equivalently, of the graph, where the coupling vertex has been removed, and the two appearing dangling bonds have been short-end terminated.}
A more rigorous foundation of this qualitative argument can be found in the Appendix.

As an illustration Fig.~\ref{fig:fig2} shows the spectrum of a tetrahedral graph.
Now $|\mathrm{Im}(G)|$ is plotted, which, up to broadening of the resonances by absorption and up to a factor, is just the density of states of the Neumann resonances.
The $G$ has been obtained from the measured reflection as described above.
Furthermore, $|\mathrm{Im}(G^{-1})|$ is shown, mirrored at the abscissa, converting the zeros of $G$ into broadened $\delta$ peaks.
The combs in the upper and the lower part of the figure mark the positions of the calculated eigenfrequencies of the closed tetrahedron with Neumann and with Dirichlet boundary conditions at the coupling vertex, respectively.

The bond lengths have been determined from transmission measurements for each individual bond, with errors, however, of several millimeters resulting from uncertainties due to the connecting junctions.
Therefore, the lengths entering the calculation have been optimized (within these uncertainties) in a least-squares fit procedure to adjust the experimental spectra to the theoretical spectra.
If this is done, perfect agreement is found (see Fig.~\ref{fig:fig2}).

The eigenvalues of the two spectra are strictly alternating, as it is the case for the toy graph as well (see Fig.~\ref{fig:fig1}).
This is a manifestation of the Neumann-Dirichlet interlacing theorem  (see Chap.~3.11 of Ref.~\cite{ber13}; for a tutorial introduction see Ref.~\cite{ber17}): 
{\em If the boundary conditions at one vertex of a graph are changed from Neumann to Dirichlet, the eigenvalues of the original and the new graph appear strictly alternating.}
In its general form the interlacing theorem allows for arbitrary changes of mixed boundary conditions between Neumann and Dirichlet, which, however, is not of relevance in the present context.

\subsection{The Neumann-Dirichlet interlacing}

There is a spectral interlacing also for $|h|$, just as for $G$ for the one-channel case.
The situation now is somewhat different however.
To move from the Neumann spectrum to the Dirichlet, one has to change the boundary conditions from Neumann to Dirichlet one after the other at {\em all} vertices, not just at one of them.
Now there is no longer a strict alternation in the sequence of the respective eigenvalues, but a strong correlation still remains:
The maximum number of Neumann eigenvalues confined between two successive Dirichlet ones is given by the number of vertices.
We checked this numerically for the tetrahedron shown in Fig.~\ref{fig:fig2} and found for a total of 942 Neumann eigenvalues, that in 38.4\%, 32.1\%, 18.5\%, 8.5\%, and 0.4\% of all cases zero, one, two, three, and four Neumann eigenvalues, respectively, were confined between two neighboring Dirichlet eigenvalues.
There was no example with more than four Neumann eigenvalues confined between Dirichlet eigenvalues, in accordance with the interlacing theorem.

\begin{figure}
	\includegraphics[width=\columnwidth]{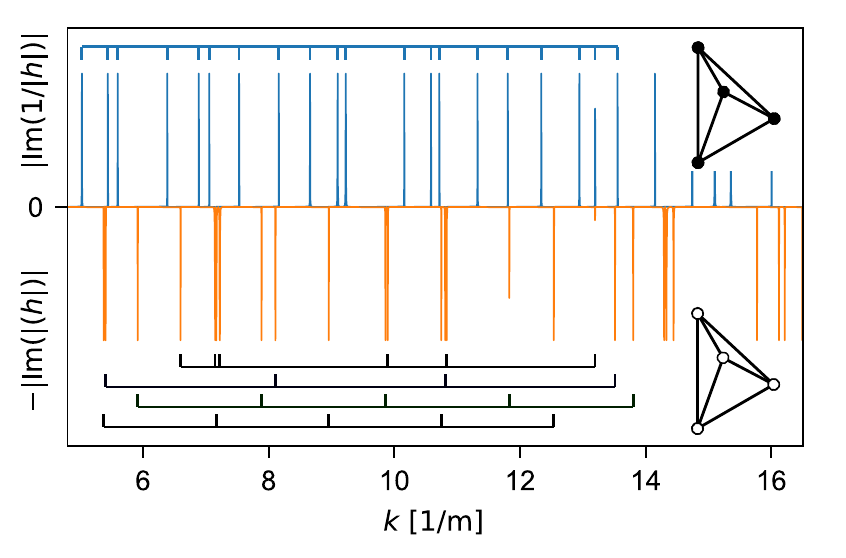}
	\caption{\label{fig:fig3}
		Numerical $|\mathrm{Im}(|h|^{-1})|$ (blue) and $-|\mathrm{Im}(|h|)|$ (orange) for the upper tetrahedral graph shown on the right.
		The lengths are the same as in Fig.~\ref{fig:fig2}.
		The combs denote the calculated spectra for the two tetrahedrons depicted on the right (see the text for details).}
\end{figure}

To determine the total $h$ matrix experimentally, one would have to attach open channels to each of the $V$ vertices and would have to measure the total $V\times V$ scattering matrix.
We spared ourselves this considerable effort and resorted to numerics.
Figure~\ref{fig:fig3} shows the results for a tetrahedron with the same lengths as in the experiment, but with the coupling vertex removed.
The upper part shows the spectrum of the Neumann resonances, obtained by adding a small imaginary part $i\varepsilon$ to $k$ and taking the imaginary part of $|h|^{-1}$.
For the purpose of a better visualization of the resonances, we did not perform the limit $\varepsilon\to 0$ but kept a non-zero value for $\varepsilon$.
The upper comb shows the same spectrum again.
Comparison with Fig.~\ref{fig:fig2} shows that the spectra of the two graphs depicted in the upper right of Figs~\ref{fig:fig2} and \ref{fig:fig3} are identical, illustrating the above-mentioned fact that the spectrum of a graph is not changed if a vertex with Neumann boundary condition is added along a bond.

The lower part of Fig.~\ref{fig:fig3} shows $|\mathrm{Im}(|h|)|$, mirrored at the abscissa, corresponding to the Dirichlet spectrum.
The combs in the lower part of the figure mark the positions of the spectra of the individual bonds constituting the Dirichlet spectrum, the three lowermost combs for the three longest bonds; 
in the upmost comb all Dirichlet eigenvalues associated with the shorter bonds are combined.

\subsection{Number variances of Neumann and Dirichlet spectra}

Each bond of a graph contributes with a series of equally distant resonances to the Dirichlet spectrum, which eventually, for large-$k$ values, add up to a sequence with more or less randomly distributed eigenvalues.
However, since on average Dirichlet and Neumann eigenvalues have to alternate, it is unavoidable that to some extent the spectral statistics of the Dirichlet eigenvalues must leave a mark on the statistics of the regular resonances.

\begin{figure}
\includegraphics[width=\columnwidth]{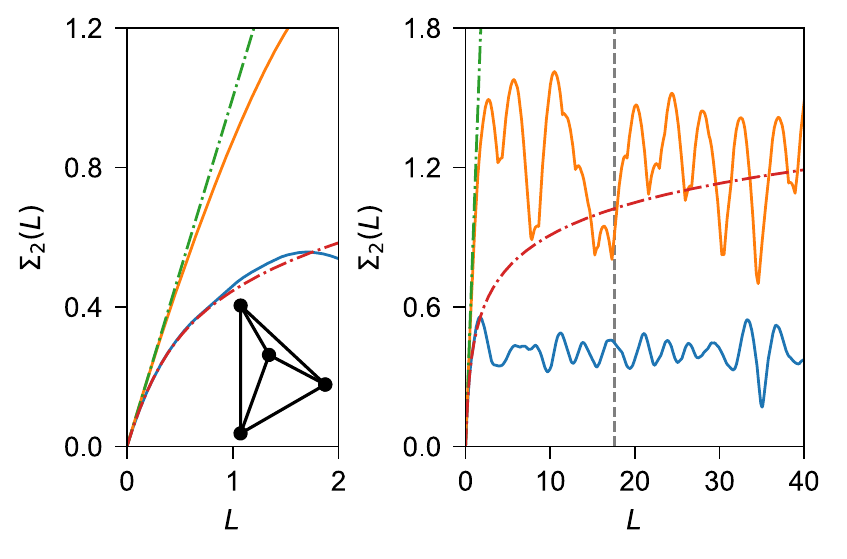}\\
\caption{\label{fig:fig4}
	Number variance $\Sigma^2(L)$ for the Dirichlet (orange) and Neumann (blue) resonances of a tetrahedral graph (numerics, lengths as in Fig.~\ref{fig:fig2}, with the coupling point removed).
	Only data for $k>\pi/l_\mathrm{min}$ have been considered.
	The dash-dotted green and red lines correspond to the expectations for the Poissonian and the Gaussian orthogonal random matrix ensemble, respectively, with the mean level spacing normalized to one.
	The vertical dashed line marks $L_\mathrm{min}=\pi/l_\mathrm{min}$, the $L$ value associated with the shortest periodic orbit.
	The left part of the figure shows the range $0<L<2$ in more detail.
}
\end{figure}

A suitable quantity to explore the consequences of this spectral interlacing is the number variance $\Sigma^2(L)$, i.e., the variance of the number of resonances in a spectral range of length $L$.
It is plotted in Fig.~\ref{fig:fig4} for both the Dirichlet and the Neumann spectrum of a tetrahedron, this time obtained from computer simulation of a closed graph.
The mean level spacing has been normalized to one, $\Delta=\pi/l_\mathrm{tot}=1$.
In addition, the expectations for a Poissonian and a Gaussian orthogonal random matrix ensemble (GOE) are shown.

In the case of many bonds with irrational lengths the Dirichlet spectrum locally approximates Poissonian statistics; 
however, the corresponding number variance does not follow the Poissonian expectation.
Responsible are the long-range spectral correlations resulting from the picket-fence structure of the spectra of the individual bonds.
The number variance for a spectrum of equidistant resonances
with a spacing of one is given by
\begin{equation} \label{eq:sigma2_top}
  \Sigma^2(L)= \{L\}\left(1-\{L\}\right)\,, \quad \{L\}=L-[L]\,.
\end{equation}
Furthermore, for different families of Dirichlet spectra $\Sigma^2(L)$ is additive as long as the lengths are incommensurable.
The curve depicted in Fig.~\ref{fig:fig4} is in agreement with this prediction with deviations of the order of the line strength.

The main message from Fig.~\ref{fig:fig4} is the number variance of the Neumann resonances.
For small values $\Sigma_2(L)$ follows the GOE expectation (see the left part of the figure), but already for $L=1.5$ it starts to deviate from the RMT expectation and eventually oscillates slowly about an average value of about 0.5.
For $L>\pi/l_\mathrm{min}$ each bond contributes with at least one resonance in a spectral range of length $L$; 
beyond this $L$ value the oscillations in the number variances of the Neumann and the Dirichlet spectrum start to synchronize, a clear indication of the correlation between the two spectra.

\begin{figure}
	\includegraphics[width=\columnwidth]{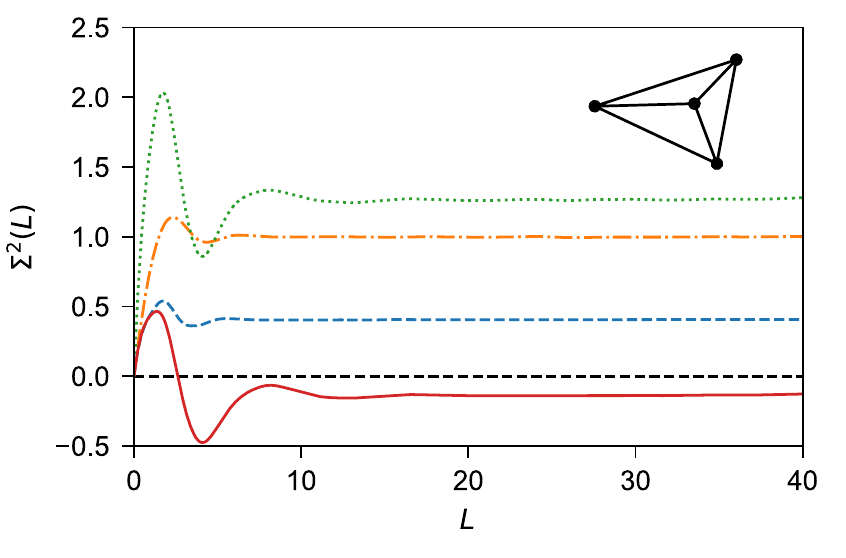}
	\caption{\label{fig:fig5}
		Number variance $\Sigma^2_{n+d}(L)$ for the sum of Neumann and Dirichlet resonances of a tetrahedral graph in a window of length $L$ (dotted green).
		To improve the statistics the results from $10^4$ tetrahedrons  of the same total length as the one shown in Fig.~\ref{fig:fig4} have been superimposed.
		In addition, the number variance of the Neumann resonances $\Sigma^2_n(L)$ (dashed blue), of the Dirichlet resonances $\Sigma^2_d(L)$ (dash-dotted orange), and the difference $\Delta_{n+d}(L)=\Sigma^2_{n+d}(L)-\Sigma^2_n(L)-\Sigma^2_d(L)$ (solid red) are shown.
	}
\end{figure}

The best tool to quantify the correlation between the two spectra is the variance of the sum (or difference) of the number of Neumann and Dirichlet resonances in an interval of length $L$, $\Sigma^2_{n+d}(L)$.
For uncorrelated Neumann and Dirichlet spectra $\Sigma^2_{n+d}(L)$ should be just the sum of number variances of the Neumann and the Dirichlet spectrum, $\Sigma^2_n(L)$ and $\Sigma^2_d(L)$, respectively.
The difference $\Delta_{n+d}(L)=\Sigma^2_{n+d}(L)-\Sigma^2_n(L)-\Sigma^2_d(L)$ is hence a measure for the correlations.
Figure~\ref{fig:fig5} shows the various number variances involved for a set of tetrahedrons of the same total length as used in the experiment, obtained by superimposing the results from $10^4$ realizations.
With increasing $L$, all number variances begin to saturate, $\Delta_{n+d}(L)$ in particular at a value of about $-0.2$, significantly different from zero, thus illustrating again the correlation between the two spectra.

\section{Discussion}

These findings require an explanation.
Hitherto there has been no doubt that the spectral statistics of graphs are well described by RMT.
There are even proofs of this fact mentioned in the Introduction.
In addition, in all experimental graph studies, including our own, agreement with RMT predictions had been found, with respect both to spectral and to scattering properties.
How does this fit together?

The standard tool to characterize spectral statistics is the spacing distribution $p(s)$ of neighboring levels.
In our experimental results we could not find any deviations from RMT predictions, but the statistical evidence was moderate with only about 800 levels involved.
The same had been done by Kottos and Smilansky \cite{kot97a} in a numerical study with a much larger data ensemble of 80\,000 levels, and in fact they {\em did} observe deviations from RMT predictions, though only on the percent level, comparable in size to the deviation between the Wigner distribution and the exact RMT expression.

This is in accordance with our own results as well as with previous results for the number variance.
Kottos and Smilansky studied $\Sigma^2(L)$ of graphs in dependence on the connectivity \cite{kot99a}.
They argued that for totally connected graphs $\Sigma^2(L)$ should approach the random matrix limit.
For $L<2$ they found similar agreement with RMT, as exhibited in the left part of Fig.~\ref{fig:fig4}; they did not show, however, results for $L>2$.
Deviations from RMT predictions for the number variance have been reported already in Refs.~\cite{die17,lu20} and for the spectral rigidity in Ref.~\cite{hul04}.
It is known from semiclassical quantum mechanics that the shortest periodic orbit shows up in a saturation of number variance and spectral rigidity \cite{ber85};
however, in the present case this cannot be the explanation.
In the graph the shortest periodic orbit is along the shortest bond with length $l_\mathrm{min}$.
The lowest resonance associated with this bond is found at $\pi/l_\mathrm{min}= l_\mathrm{tot}/l_\mathrm{min}$ [using $l_\mathrm{tot}=\pi$, following from the normalization of the mean level spacing to one (see above)].
Thus saturation is expected beyond this value, i.e., for the present graph at $L=17.6$, by far beyond $L=1.5$, where saturation starts.
Thus the saturation must have another origin, and our explanation is the interlacing of the Neumann and the Dirichlet spectrum.

This might explain why the consequences of spectral duality have remained unnoticed so far: 
The picket-fence structure of the Dirichlet spectrum leaves its marks in the long-range correlation, but only slightly influences the near-distance Neumann eigenvalue statistics.

This is not in contradiction with the proofs that the spectra of incommensurable graphs obey RMT statistics: 
The theory works in the limit of large vertex numbers, whereas in the experiments as well as in the numerical studies the vertex number typically is below 10-20.
However, even for a large number of vertices $V$ the interlacing feature of the Dirichlet and Neumann spectra is still present:
After $V$ Neumann eigenvalues at the latest a Dirichlet eigenvalue appears and vice versa.
Thus, on the $k$ axis there are windows containing up to $V$ Neumann eigenvalues alternating with windows containing up to $V$ Dirichlet eigenvalues.
Within each Dirichlet window the eigenvalues are Poisson distributed, but (and this is the essential point) a Dirichlet window contributes {\em only once} to the Neumann nearest level spacing distribution $p(s)$.
Thus, in the limit $V\to\infty$ the contribution from the Dirichlet windows to $p(s)$ becomes negligible.
This is a qualitative way to reconcile spectral interlacing with the RMT behavior of the Neumann resonances for large $V$, but it shows at the same time that essential features are missed in the RMT approach; 
it reflects only half of the truth!

\section{Conclusion}

The conclusion is clear:
For graphs the range of validity of RMT is restricted to at most $V$ neighbors, where $V$ is the number of vertices.
This does not mean that one has to question all previous graph experiments.
Many of them concentrated on level spacing statistics, which anyway is sensitive mainly to the level repulsion of close neighbors.
But whenever larger distance properties are involved, a study of the interplay of Neumann and Dirichlet eigenvalues is mandatory.
The discussion of the variance of the sum of Neumann and Dirichlet eigenvalues in a window of a given length $L$, presented in this paper, means a first step in this direction.

\begin{acknowledgments}
	We are grateful to Sven Gnutzmann, Nottingham, and Holger Schanz, Magdeburg, for clarifying discussions, and for calling our attention to a number of relevant references. J.L.~acknowledges financial support from the China Scholarship Council via Grant No.~202006180008.
\end{acknowledgments}

\section*{Appendix}

Here we proof that the zeros of $G$ [see Eq.~(\ref{eq:G2})] are the eigenvalues of the graph obtained by removing vertex $1$, the coupling vertex, and by terminating the emerging dangling bonds by Dirichlet boundary conditions, in the following shortly termed the ``truncated graph''.
For the sake of simplicity, we assume that there are just two bonds connecting the coupling vertex via vertices $L$ and $R$ to the rest of the graph, but the proof holds for an arbitrary number of coupling bonds.

Applying the sequence $1$, $L$, $R$, and then the rest of the graph, of rows and columns, the secular matrix $h$ [see Eq.~(\ref{eq:hsec})] may be written as
\begin{equation} \label{eq:A1}
h=\left(
\begin{array}{c|c}
-f_L-f_R &\quad g^T \\[0.5ex]\hline\vspace{0.5ex}
g &\quad \hat{h} \\
\end{array}
\right)\,,
\end{equation}
where
\begin{equation} \label{eq:A2}
g^T=(g_L,g_R,0,\dots, 0)
\end{equation}
and $\hat{h}$ is the secular matrix of the truncated graph.
Elementary matrix calculation yields
\begin{equation} \label{eq:A3}
G=\left(h^{-1}\right)_{11}=\left[-f_L-f_R-g^T\hat{h}^{-1} g\right]^{-1}\,.
\end{equation}
The spectrum of the truncated graph is given by the zeros of $|\hat{h}|$.
They show up in the poles of the denominator, resulting in zeros of $G$.
All eigenvalues of the truncated graph thus generate zeros of $G$.
To complete the proof we have to show that there are {\em no further} zeros of $G$, not associated with the spectrum of the truncated graph.
The only possible candidates are the poles of $f_L$ and $f_R$; however, these poles are canceled by the corresponding poles of the third term in the denominator, which we are now going to prove.

For the rest of this appendix we assume that $\hat{h}$ is invertible, i.e., we avoid the positions of the resonances of the truncated graph.
Similarly to the above we write $\hat{h}$ in block form,
\begin{equation} \label{eq:A4}
\hat{h}=\left(
\begin{array}{c|c}
-F-\tilde{F} &\quad \tilde{g}^T \\[0.5ex]\hline\vspace{0.5ex}
\tilde{g} &\quad \tilde{h} \\
\end{array}
\right)\,.
\end{equation}
The upper left corner element contains the $L,R$-block with
\begin{equation} \label{eq:A5}
F=\left(
\begin{array}{cc}
f_L & 0 \\
0 & f_R \\
\end{array}
\right)\,,\quad \tilde{F}=\left(
\begin{array}{cc}
\sum f_{Li} & -g_{LR} \\
-g_{LR} & \sum f_{Ri} \\
\end{array}
\right)\,,
\end{equation}
where the sums $\sum f_{Li}$ and $\sum f_{Ri}$ run over all bonds connecting vertices $L$ and $R$, respectively, with the rest of the graph. The $g_{LR}$ term is present only, if vertices $L$ and $R$ are directly connected via a bond. Here $\tilde{h}$ is the secular matrix of the graph obtained by removing vertices $L$, and $R$, and terminating the emerging dangling bonds by short ends, and $\tilde{g}$ is the matrix containing the $g_{Li}$ and $g_{Ri}$ terms of the coupling vertices $L$ and $R$, respectively, to the rest of the graph.
It follows that
\begin{equation} \label{eq:A6}
g^T \hat{h}^{-1} g = (g_L g_R)\left[-F-\tilde{F}-\tilde{g}^T\tilde{h}^{-1}\tilde{g}\right]^{-1}\left(
\begin{array}{c}
g_L \\
g_R \\
\end{array}
\right)
\end{equation}
Using $g_n^2-f_n^2=1$, $n=L,R$, following immediately from the definitions~(\ref{eq:fg}), we obtain
\begin{equation} \label{eq:A7}
-f_L-f_R= \Tr F^{-1}-(g_L g_R)F^{-1}\left(
\begin{array}{c}
g_L \\
g_R \\
\end{array}
\right)\,.
\end{equation}
Plugging the expressions (\ref{eq:A6}) and (\ref{eq:A7}) into Eq.~(\ref{eq:A3}), we obtain
\begin{equation} \label{eq:A8}
G=\left[\Tr F^{-1}+(g_L g_R)D\left(
\begin{array}{c}
g_L \\
g_R \\
\end{array}
\right)\right]^{-1}\,,
\end{equation}
where
\begin{eqnarray} \label{eq:A9}
\nonumber
D &=&\left[F+\tilde{F}+\tilde{g}^T\tilde{h}^{-1}\tilde{g}\right]^{-1}-F^{-1}\\\nonumber
&=&\left[F+\tilde{F}+\tilde{g}^T\tilde{h}^{-1}\tilde{g}\right]^{-1}\\\nonumber
&&\times\left[F-(F+\tilde{F}+\tilde{g}^T\tilde{h}^{-1}\tilde{g})\right] F^{-1}\\
&=&-\left(F+\tilde{F}+\tilde{g}^T\tilde{h}^{-1}\tilde{g}\right)^{-1}\!\!\!(\tilde{F}+\tilde{g}^T\tilde{h}^{-1}\tilde{g})\,F^{-1}.
\end{eqnarray}
This was the crucial step; the singularities resulting from the $f_L$ and $f_R$ dropped out.

Inserting the result into Eq.~(\ref{eq:A8}), we get
\begin{equation} \label{eq:A10}
G=\left[\Tr F^{-1}-(g_L g_R)F^{-1}\tilde{D}F^{-1}\left(
\begin{array}{c}
g_L \\
g_R \\
\end{array}
\right)\right]^{-1}\,,
\end{equation}
where 
\begin{eqnarray} \label{eq:A11}
\nonumber
\tilde{D} &=&\!-FDF\\
&=&\!\left(1+\tilde{F}F^{-1}+\tilde{g}^T\tilde{h}^{-1}\tilde{g}F^{-1}\right)^{-1}
\!\!\!(\tilde{F}+\tilde{g}^T\tilde{h}^{-1}\tilde{g})\,.
\end{eqnarray}
Now the limits $\sin kl_L \to 0 $ or $\sin kl_R \to 0$ can be performed, with all terms depending on $kl_L$ or $kl_R$, and $f_n^{-1}=\tan kl_n$ and $g_n/f_n=1/\cos kl_n$ ($n=L,R$), remain regular. Q.\,E.\,D.

\end{document}